\documentclass[a4paper,11pt]{article}
\usepackage{pos}

\usepackage{natbib}  
\title{Convolution and Graph-based Deep Learning Approaches for Gamma/Hadron Separation in Imaging Atmospheric Cherenkov Telescopes}
\ShortTitle{Convolution and graph-based deep learning approaches for $\gamma$/hadron separation in IACTs}

\author*[a,b]{Abhay Mehta}
\author[a,b]{Dan Parsons}
\author[a]{Tim Lukas Holch}
\author[a,b]{David Berge}
\author[b]{Matthias Weidlich}

\affiliation[a]{Deutsches Elektronen-Synchrotron DESY, \\ Platanenallee 6, 15738 Zeuthen, Germany}

\affiliation[b]{Humboldt-Universit{\"a}t zu Berlin,\\ Unter den Linden 6, 10117 Berlin, Germany}

\emailAdd{abhay.mehta@desy.de}

\abstract{The identification of $\gamma$-rays from the predominant hadronic-background is a key aspect in their ground-based detection using Imaging Atmospheric Cherenkov Telescopes (IACTs). While current methods are limited in their ability to exploit correlations in complex data, deep learning-based models offer a promising alternative by directly leveraging image-level information. However, several challenges involving the robustness and applicability of such models remain. Designing model architectures with inductive biases relevant for the task can help mitigate the problem. Three such deep learning-based models are proposed, trained, and evaluated on simulated data: (1) a hybrid convolutional and graph neural network model (CNN-GNN) using both image and graph data; (2) an enhanced CNN-GNN variant that incorporates additional reconstructed information within the graph construction; and (3) a graph neural network (GNN) model using image moments serving as a baseline. The new combined convolution and graph-based approach demonstrates improved performance over traditional methods, and the inclusion of reconstructed information offers further potential in generalization capabilities on real observational data.

}

\ConferenceLogo{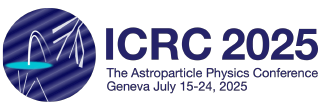}

\FullConference{39th International Cosmic Ray Conference (ICRC2025)\\
 15–24 July 2025\\
Geneva, Switzerland\\}

\begin{document}
\maketitle

\section{Introduction}
The field of very-high-energy (VHE) $\gamma$-ray astronomy has evolved significantly over the past three decades, driven largely due to observations from ground-based imaging telescopes. These Imaging Atmospheric Cherenkov Telescopes (IACTs) capture Cherenkov light produced by highly energetic particles interacting with the Earth’s atmosphere \cite{deNaurois_2015}. Each camera image represents a two-dimensional projection of the Cherenkov light pool from the telescope's position, and each event typically consists of multiple images taken simultaneously across an array of telescopes. This stereoscopic information is used to reconstruct the energy, direction, and type of the primary particle initiating the air shower. Event statistics, however, are dominated by hadron-induced air showers, which can outnumber $\gamma$-ray air showers by up to a factor of $10^4$ \cite{HESS_2006}. This makes the task of identifying $\gamma$-ray events from the hadronic background a central challenge for IACT-based observations.

Hadronic air showers are fundamentally different from those initiated by $\gamma$-rays, which are electromagnetic in nature. This difference in shower development leads to small but significant differences in the camera images, which form the basis for separating the two event classes. Current-generation IACTs typically rely on Boosted Decision Trees (BDTs) trained on parameterized image features or goodness-of-fit parameters for this task \cite{Hillas_1985, Ohm_2009, deNaurois_2009}. Consequently, a natural motivation for exploring deep learning-based models stems from the possibility of improving event classification by directly using image-level information.

\section{Deep Learning-based Classifiers for IACTs}
Multiple studies have explored deep learning methods for identifying $\gamma$-rays and demonstrated exceptional performance on simulated data \cite{Shilon_2019, Parsons_2020, Spencer_2021, Nieto_2019, Jacquemont_2020}. Most model architectures use convolutional neural networks (CNNs) for extracting information from camera images and recurrent neural networks (RNNs) for its aggregation across an event. Graph neural networks (GNNs), applied on images represented as point clouds, have also been established as a viable approach for the same task \cite{Glombitza_2023}. Despite their potential, a complete deployment of deep learning-based models on IACT data remains non-trivial due to a variety of issues, ranging from observational systematics to discrepancies between simulations and real-world data. 



The construction of models that can generalize to ``unseen" situations is a long-standing problem in deep learning \cite{Mitchell_1980}. The use of network architectures with inductive biases suitable for the given task can lead to improved generalizations by guiding the learning process towards more physically meaningful representations \cite{Kipf_2020, google_inductive_2018}. For example, translational invariance in CNNs, temporal dependence in RNNs, and permutation equivariance in GNNs are all properties that align naturally with the structure of the data these models are typically applied to, thus contributing to their success in their domains. In the context of IACTs, as each image is a projection of the \emph{same} event, there is no inherent ordering between them, making them permutation equivariant. This motivates a shift towards exploring GNNs for aggregating information across multiple telescopes. To that end, this work introduces a combined convolutional and graph neural network (CNN-GNN) based approach for $\gamma$/hadron separation in IACTs. 

Three models, with two distinct training strategies, are proposed and evaluated on simulated data. These include a CNN-GNN model trained on image and graph data, an enhanced CNN-GNN variant with additional reconstructed event information incorporated into the graph structure, and a baseline GNN model utilizing image moments, serving as a reference for existing methods.





\section{Datasets and Input Processing} \label{sec:datasets}
The models were trained on simulations of the High Energy Stereoscopic System (H.E.S.S.) located in Göllschau, Namibia \cite{HESS_2006}. The H.E.S.S. array consists of four 12-meter telescopes (CT1-4) arranged in a square of side 120 meters, with an additional 28-meter telescope (CT5) at its center. Diffuse proton and $\gamma$-ray events, simulated at a zenith of $20^{\circ}$ and with a maximum view cone of $5^{\circ}$, were chosen as the two event classes for this task. Similar models were also trained on the simulations of the upcoming Cherenkov Telescope Array Observatory \cite{CTAO_2017}, the results of which are not presented here. 

A custom data processing pipeline, based on \texttt{ctapipe} and \texttt{PyTorch}, was developed for the subsequent analysis of IACT data. The framework within the \texttt{ctapipe} v0.19.3 \cite{ctapipe} package was used to handle the low-level data processing tasks such as image calibration and cleaning, while the construction and training of the convolution and graph-based models were implemented using \texttt{PyTorch} v2.0.1 \cite{pytorch} and \texttt{PyTorch Geometric} v2.4.0 \cite{pytorch_geometric}, respectively.

\begin{figure}[h]
    \centering
    \includegraphics[width=\linewidth]{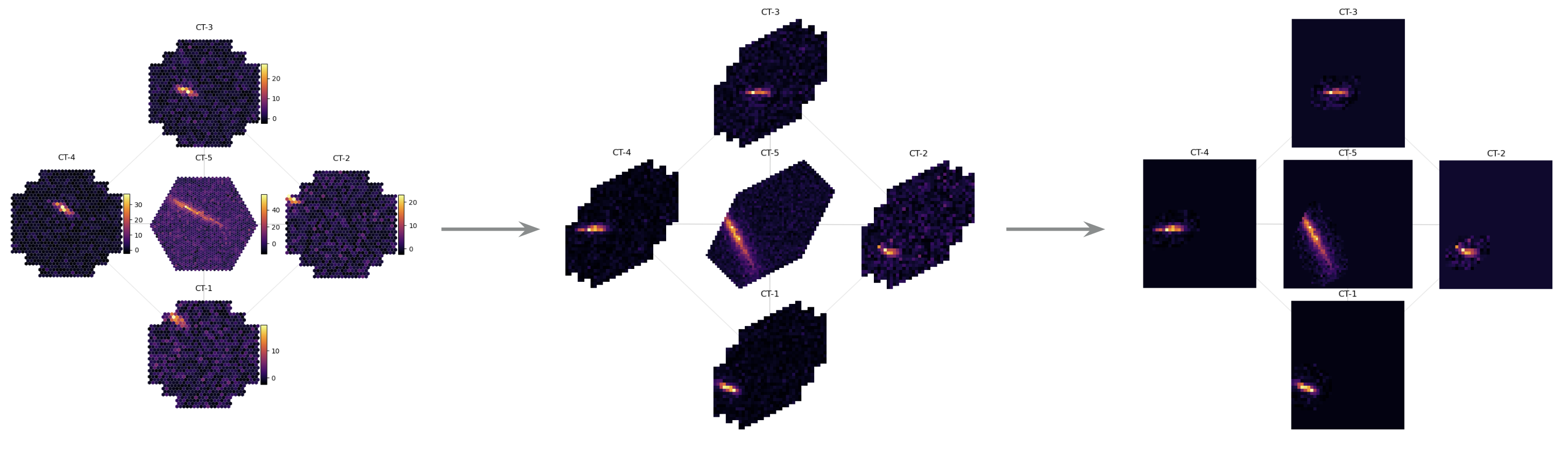}
    \caption{Schematic visualization of the  image processing pipeline of images in an event for input into a convolutional neural network.}
    \label{fig:Data_process_image}
\end{figure}
All images were first cleaned with a standard tailcuts procedure, while preserving information from four neighboring layers of surviving pixels. The cleaned hexagonal-pixel images were then re-binned into a square-pixel format for compatibility with convolutional architectures (Figure \ref{fig:Data_process_image}). A minimum intensity cut was applied on each image, and the resulting images were normalized by their maximum pixel intensity. Only events with at least two non-empty images after this step were retained for training. Fully connected graphs, with each telescope as a node, were constructed for each event. Node-level information included, along with the images, the maximum and total pixel intensity, positions of the telescope in ground reference frame, and the reconstructed impact position of the primary particle on the ground. The final dataset consisted of 150\,000 events, evenly split between $\gamma$-ray and proton events, with $10\%$ kept for validation and another $10\%$ for testing. 

A secondary graph dataset was created using only the reconstructed parameterized information (i.e. Hillas parameters \cite{Hillas_1985}) of the same events, to serve as an input for a baseline model.

\newpage
\section{Models and Training Strategies}
The primary objective of the models developed in this work is to classify events as either $\gamma$-ray or hadron-induced based on the event images and information. To this end, the aim of the convolutional (CNN) half of the model remains to extract relevant image-level features from each telescope, while the graph-based (GNN) half is given additional contextual information (described in Section \ref{sec:datasets}) to better learn the classification task. The output of the CNN also forms part of the input to the GNN, allowing the model to combine image features with telescope-level information during training.

A key advantage of the graph representation in GNNs is its flexibility in handling varying numbers and types of telescopes present in IACT arrays \cite{deNaurois_2015}. This is also relevant for when subsets of the array operate due to technical or observational constraints. Additionally, the graph approach also allows for inclusion of telescope-specific information, such as position, reconstructed parameters, temporal relationships, making the system more adaptable to the particularities of real-world data. 

The CNN component is a simple implementation inspired from the Inception module \cite{inception_paper}, employing convolution filters of multiple sizes in parallel to learn features at different spatial scales. Five convolutions with kernel sizes ranging from 1 to 15 are applied to each image and the output subsequently passed through two additional convolutional layers with max-pooling. The result is then flattened and processed through two fully connected layers, with dropout regularization applied before the final layer. The network operates on images from each telescope independently and returns a fixed-dimensional embedding for each telescope.

The feature vectors extracted by the CNN are subsequently passed to the GNN component, which consists of a multi-layer EdgeConv architecture with residual concatenation across layers, following the original design \cite{EdgeConv_2019}. After message passing, the learned node representations are aggregated globally using a combination of mean, max, and sum pooling operations to form a graph-level event embedding, which is then used for classification. Alternative GNN layers such as GCNConv and GATConv were also explored, but no significant difference in performance was observed.

\begin{figure}[h]
    \centering
    \includegraphics[width=\linewidth]{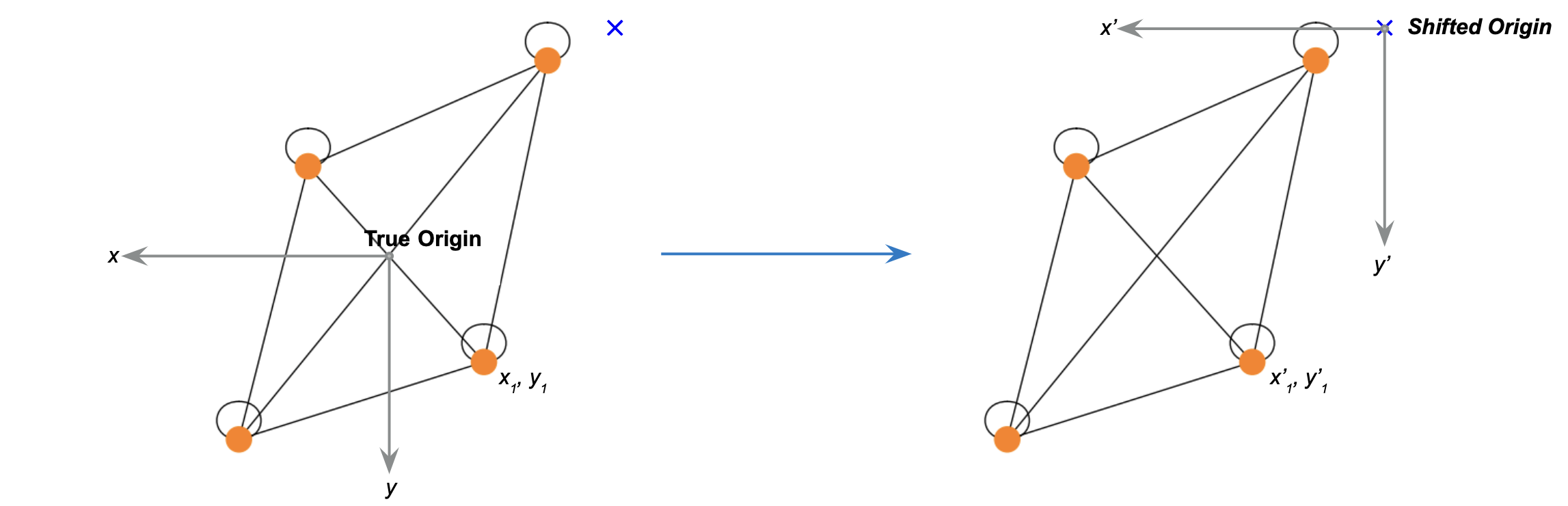}
    \caption{Sketch illustrating the shift in coordinate frames with the new origin centered at the reconstructed core position of an event. The orange nodes indicate four IACT telescopes with the black lines indicating the edges of the graph structure. The blue cross marks the reconstructed impact point of the primary particle. This transformation is used for the inclusion of reconstructed information in the Strong CNN-GNN approach.}
    \label{fig:shift_coords}
\end{figure}

A key aim of the work was to examine if the inclusion of reconstructed information in the model training contributed to improvements in performance. Consequently, three models with different GNN components were developed: a Fast CNN-GNN, a Strong GNN and a Hillas GNN. The specifics follow-
\begin{itemize}
    \item \textbf{Fast} CNN-GNN: Combines features extracted from camera images with telescope positions (in the ground frame), total image and maximum pixel intensity per image. 
    \item \textbf{Strong} CNN-GNN: An enhanced version which incorporates the reconstructed core position of the air shower. This is implemented by shifting the telescope positions in the tilted ground frame\footnote{The tilted ground frame is the ground frame transformed according to the telescope's pointing direction and is used for reconstructing the shower core position.} by centering the origin at the reconstructed impact point (Figure \ref{fig:shift_coords}). 
    \item \textbf{Hillas} GNN: A simplified GNN trained solely on Hillas parameters extracted from the camera images serving as the baseline model.
\end{itemize}

All models were trained as binary classifiers to distinguish between $\gamma$-ray and proton events, using the \texttt{BinaryCrossEntropy} loss function. The output of each model is a $\Gamma$-score in the range $[0,1]$, where values closer to 0 and 1 correspond to proton-like and $\gamma$-ray–like events, respectively. Models were trained up to 51 epochs, although early stopping was used around epoch 20–25. The Adam optimizer was used with an initial learning rate of $10^{-3}$ and a batch size of 72. A \texttt{ReduceLROnPlateau} scheduler was employed to adaptively lower the learning rate upon stagnation in validation accuracy, although this had limited impact in most cases. \\

The standard training strategy of \textit{end-to-end} training, where the weights of both the CNN and GNN components are free and optimized together, was used. This was tested within a homogeneous graph setting for which CT5 was excluded, but an extension to a heterogeneous framework to include other types of telescopes is easily feasible. The approach allows the model to effectively learn and exploit correlations across the full event. A drawback of this approach is the lack of interpretability of the features extracted by the CNN.

An alternative approach explored was that of \textit{split} training, in which the two components are trained separately. First, the CNN is trained on single-telescope images for each camera type to produce a CNN-score. These outputs, together with other telescope-specific information, are then used to construct a graph representation of each event. The GNN is subsequently trained on this graph data, with the CNN weights held fixed. This strategy reduces the risk of simulation overfitting and retains greater interpretability of the learned features. However, such an approach leads to a trade-off between robustness and performance, since it is less effective at learning event representations. On simulated data, the split training approach showed limited performance gains compared to existing methods for this task.


\section{Performance}

\begin{figure}[h]
    \centering
    \includegraphics[width=0.7\linewidth]{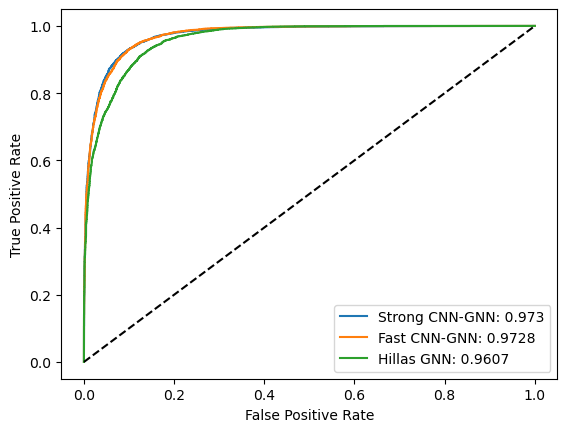}
    \caption{Receiver Operating Characteristic (ROC) curves for the three models, with the Area Under Curve (AUC) values indicated in the legend. }
    \label{fig:roc_plot}
\end{figure}

All models were tested on a testing dataset described in Section \ref{sec:datasets}. The Receiver Operating Characteristic (ROC) curves for the three models are shown in Figure \ref{fig:roc_plot}, with the values in the legend corresponding to the area under curve (AUC) values in each case. All models are successfully able to learn to classify events between $\gamma$-rays and protons and the CNN-GNN models outperform the Hillas-based GNN model as well. Upon testing on a dataset with an additional local distance cut on images, as in \cite{Shilon_2019}, the CNN-GNN approach matches the classification performance of the previous models. Similar results were also seen when the same approach was applied and evaluated on simulations of the CTAO array.

Only a small improvement in performance is seen between the two convolution and graph-based approaches in the ROC curves. However, the Strong CNN-GNN demonstrated a greater ability to reject proton events over the Fast CNN-GNN. At a decision threshold of 0.5, this corresponded to an additional 9\% reduction in the False Positive Rate relative to the Fast variant. The true impact of this approach can only be realized when evaluating on real observational data, which is heavily background dominated. Nevertheless, the addition of reconstructed information appears to have a positive effect on the classification task. As training datasets are expanded to include events from a wider range of pointings, including additional contextual information, such as zenith (or even azimuth), will strengthen the classification and generalizing capabilities of this approach.

\begin{figure}[h]
    \centering
    \includegraphics[width=0.45\linewidth]{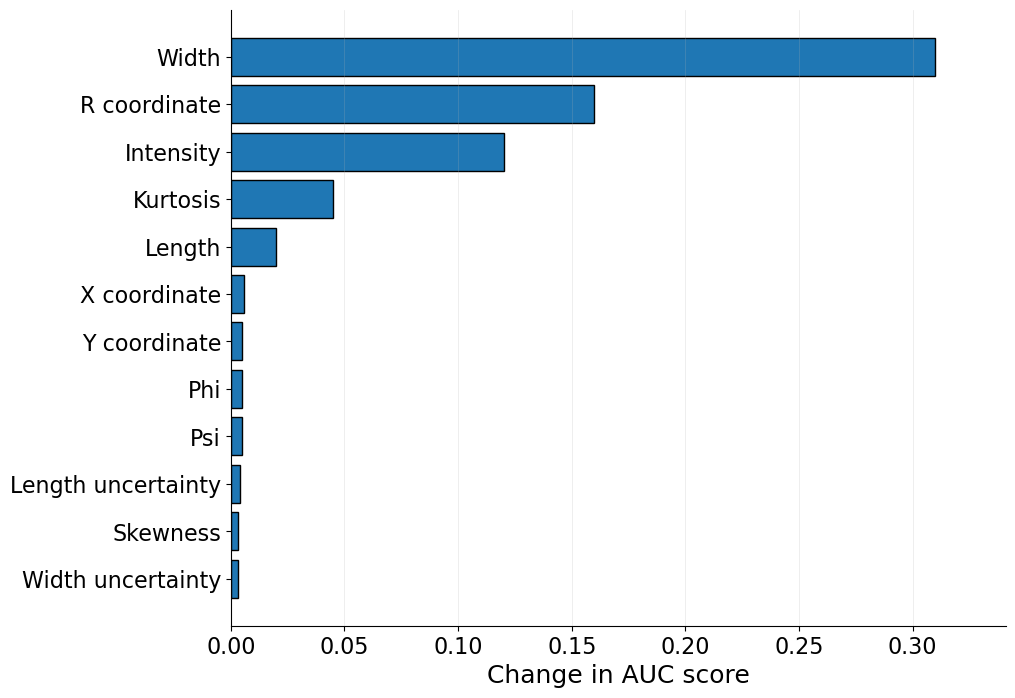}
    \includegraphics[width=0.45\linewidth]{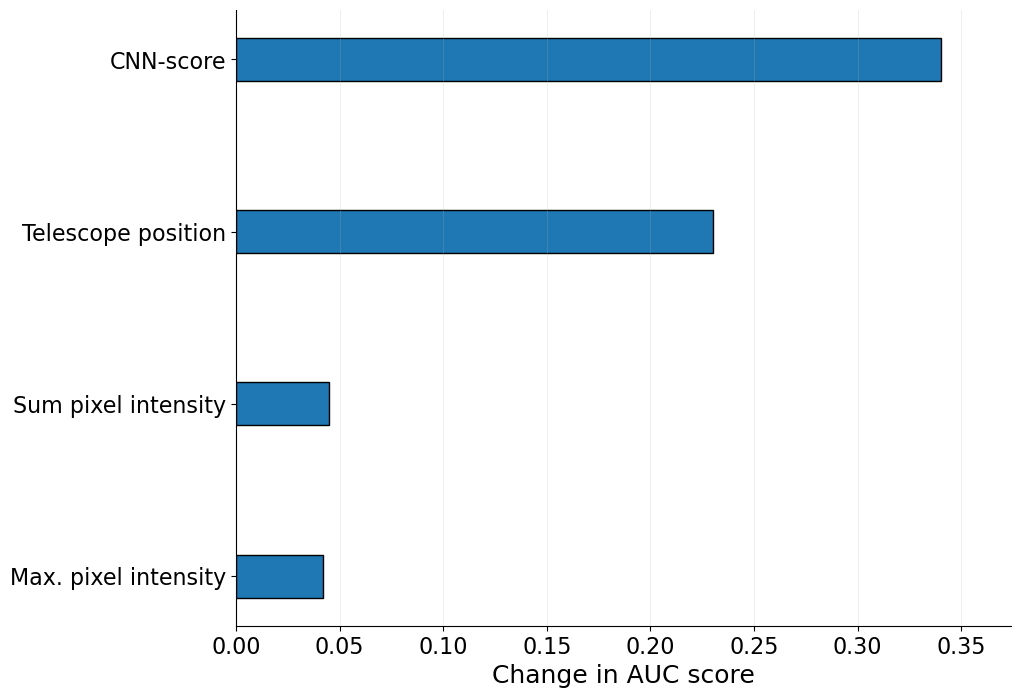}

    \caption{Permutation Feature Importance scores for the Hillas GNN and Fast CNN-GNN (split-training) models. A higher value indicates greater importance of that feature in the model's performance.}
    \label{fig:pfi_plot}
\end{figure}

To get a reliable estimate of the importance of the model features in the classification task, permutation feature importance scores were computed for select models. This computation quantifies each feature's contribution to the model's discriminative power. Here, this was done by independently shuffling each feature's values across the dataset and then measuring the change in the AUC score of the same model. The importance of a feature $f$ can then be defined as:
\begin{equation*}
\Delta \text{ AUC}_f = \text{True AUC} - \frac{1}{n} \sum_{}^{n} \text{Shuffled AUC}_f
\end{equation*}
where a higher $\Delta \text{ AUC}_f$ score indicates a greater importance of that feature to the model's decision process. The results of this computation for the Hillas GNN ($n=50$) and the Fast CNN-GNN ($n=10$) model trained via the split-training approach are shown in Figure \ref{fig:pfi_plot}. Note that these scores are not unique and depend on the specific instance of the trained model. Even with the same training data, different models can assign varying importance to each feature based on how they converged during the learning process for the classification task.

\section{Conclusions}
The applicability of deep learning-based models on IACT observational data remains a key challenge for several reasons. The observational conditions under which such data is taken are far more diverse than those that can be simulated. Additionally, physical uncertainties introduce further discrepancies between simulated and actual observations. Consequently, deep learning models are challenged in their ability to generalize given such conditions.

This work addresses these issues by exploring model architectures that include inductive biases that are aligned with the physical structure of IACT data. Convolution and graph-based methods are therefore explored for the classification of events between $\gamma$-rays and the hadronic background. The proposed hybrid CNN-GNN models show improvements in classification over parametrized approaches and match the performance of previous deep learning models, thus establishing the technique as a viable approach. Furthermore, the inclusion of reconstructed information within the graph representation seems beneficial and offers the potential to enhance the performance and generalization of such models. 

The natural next step is the evaluation of these models on real observations, which will offer a more definitive test of the viability and robustness of these approaches with real-world IACT data. This evaluation will also provide greater insight into the extent to which inductive bias\mbox{-}driven architectures can bridge the gap between simulations and reality, and contribute to the next generation of $\gamma$/hadron separation strategies.

\section{Acknowledgments}
The authors thank the H.E.S.S. collaboration for the access to and use of the H.E.S.S. simulations. They also thank Iftach Sadeh for the valuable input and feedback during the course of this work.

\bibliography{bib}
\end{document}